\documentclass[12pt]{article}
\usepackage{amsmath,amsthm,amssymb,euscript,epsfig,youngtab}
\usepackage{array}
\usepackage{fancybox}
\usepackage[usenames]{color}
\usepackage{amsfonts,bm}
\usepackage{graphicx}

\newcommand{\be}{\begin{equation}}
\newcommand{\ee}{\end{equation}}
\newcommand{\bea}{\begin{eqnarray}\displaystyle}
\newcommand{\eea}{\end{eqnarray}}

\makeatletter
\@addtoreset{equation}{section}
\makeatother
\renewcommand{\theequation}{\thesection.\arabic{equation}}
\def\one{{\hbox{ 1\kern-.8mm l}}}
\def\zero{{\hbox{ 0\kern-1.5mm 0}}}

\begin{document}

\makeatletter
\@addtoreset{equation}{section}
\makeatother
\renewcommand{\theequation}{\thesection.\arabic{equation}}

\rightline{WITS-CTP-097}
\vspace{1.8truecm}

\vspace{15pt}

%%%%%%%%%%%%%%%%%

{\LARGE{   
\centerline{\bf Nonplanar integrability at two loops} 
}}  

\vskip.5cm 

\thispagestyle{empty} \centerline{
    {\large \bf Robert de Mello Koch\footnote{{\tt robert@neo.phys.wits.ac.za}}, 
                Garreth Kemp\footnote{{\tt Garreth.Kemp@students.wits.ac.za}},
                Badr Awad Elseid Mohammed\footnote{{\tt Badr.Mohammed@students.wits.ac.za}},
                }}

\centerline{{\large \bf and Stephanie Smith\footnote{{\tt Stephanie.Smith@students.wits.ac.za}}}
                                                       }

\vspace{.4cm}
\centerline{{\it National Institute for Theoretical Physics ,}}
\centerline{{\it Department of Physics and Centre for Theoretical Physics }}
\centerline{{\it University of Witwatersrand, Wits, 2050, } }
\centerline{{\it South Africa } }

\vspace{1.4truecm}

%%%%%%%%%%%%%%%%%
\thispagestyle{empty}

\centerline{\bf ABSTRACT}

\vskip.4cm 

In this article we compute the action of the two loop dilatation operator on restricted Schur polynomials
that belong to the su$(2)$ sector, in the displaced corners approximation. 
In this non-planar large $N$ limit, operators that diagonalize the one loop dilatation operator are not corrected at two loops. 
The resulting spectrum of anomalous dimensions is related to a set of decoupled harmonic oscillators, indicating integrability 
in this sector of the theory at two loops. 
The anomalous dimensions are a non-trivial function of the 't Hooft coupling, with a spectrum that is continuous and starting
at zero at large $N$, but discrete at finite $N$.

\setcounter{page}{0}
\setcounter{tocdepth}{2}

\newpage

\tableofcontents

\setcounter{footnote}{0}

\linespread{1.1}
\parskip 4pt

\section{Introduction and Questions}

The discovery of a rich integrable structure\cite{Minahan:2002ve,Beisert:2003tq} underlying the planar limit of ${\cal N} = 4$ 
super Yang-Mills theory is fascinating.
It has allowed tremendous progress in exploring planar ${\cal N}=4$ super Yang-Mills theory and free IIB superstrings on the 
AdS$_5\times$S$^5$ background, providing novel support for the AdS/CFT 
correspondence\cite{Maldacena:1997re,Witten:1998qj,Gubser:1998bc}. 
Indeed, there is reason to hope that the exact spectrum of anomalous dimensions can be found in the planar limit.
We refer the reader to \cite{Beisert:2010jr} for a comprehensive recent review.
Even more recently, attempts to compute the three point functions of the theory using integrability have 
begun\cite{Escobedo:2010xs,Escobedo:2011xw,Gromov:2011jh,Foda:2011rr,Gromov:2012vu,Kostov:2012jr}.
An optimist might hope that planar ${\cal N}=4$ super Yang-Mills theory can be solved exactly.

A natural question to ask now, is if integrability is present beyond the planar limit.
In this article we will study the large $N$ limit of the anomalous dimensions of a class of operators, restricted
Schur polynomials\cite{Balasubramanian:2004nb,Bhattacharyya:2008rb}, that have classical dimension of order $N$.
For these operators, summing the planar diagrams does not capture the large $N$ limit\cite{Balasubramanian:2001nh}.

There has by now been some progress in the study of these highly non-trivial large $N$ limits.
The basic new ingredient has been to use the representation theory of symmetric and unitary groups.
In particular, the problem of diagonalizing the free field inner product for single and multi-matrix operators, 
to all orders in $1/N$, has been solved.
In the half-BPS sector a complete set of operators is given by the Schur polynomials $\chi_R(Z)$\cite{cjr}.
They are labeled with Young diagrams $R$. 
Operators with $R$ having order one rows of length order $N$ or order one columns of length order $N$ are dual to
giant gravitons\cite{mst,myers,hash}.
If $R$ has $O(N^2)$ boxes the corresponding operator is dual to an LLM geometry \cite{Lin:2004nb}.
The problem of diagonalizing the free field inner product for multi-matrix operators, while preserving 
global symmetries, was solved in \cite{BHR1,BHR2}.
A basis relevant for the description of brane-antibrane systems was given in \cite{Kimura:2007wy,Kimura:2009wy}.
Finally, the restricted Schur basis was proved to diagonalize the inner product in \cite{Bhattacharyya:2008rb} 
and to be complete in \cite{Bhattacharyya:2008rc}.

In this article our focus is on restricted Schur polynomials and on the su$(2)$ sector of the theory.
In the su$(2)$ sector one considers a restricted Schur polynomial built mainly from one type of matrix field $Z$,
doped with impurities $Y$. 
The restricted Schur polynomial is given by
\bea
  \chi_{R,(r,s)\alpha\beta}(Z,Y)={1\over n! m!}\sum_{\sigma\in S_{n+m}}{\rm Tr}_{(r,s)\alpha\beta}\left(\Gamma^{(R)}(\sigma)\right)
                   {\rm Tr}(\sigma Z^{\otimes n}\otimes Y^{\otimes m})
  \label{restschur}
\eea
The polynomial above is built using $n$ $Z$s and $m$ $Y$s. 
$R\vdash m+n$ is an irreducible representation (irrep) of $S_{n+m}$ and $\Gamma^{(R)}(\sigma)$ is a 
matrix representing $\sigma$ in this irrep $R$. 
Spelling out the structure of the second trace
\bea
 {\rm Tr}(\sigma Z^{\otimes n}\otimes Y^{\otimes m})= Y^{i_1}_{i_{\sigma(1)}}\cdots Y^{i_{n}}_{i_{\sigma(n)}}
                                       Z^{i_{n+1}}_{i_{\sigma(n+1)}}\cdots Z^{i_{n+m}}_{i_{\sigma(n+m)}}
\eea
we see that $\sigma$ acts on $n+m$ indices (we think of each index as a ``slot'' that can be populated with a matrix)
with the first $m$ slots associated to $Y$s and the next $n$ slots associated to $Z$s. 
Consider the $S_n\times S_m$ subgroup that permutes the $Z$ and $Y$ slots separately.
The irrep $R\vdash m+n$ will, in general, subduce many irreps of this subgroup.
Denote these irreps by $(r,s)\alpha$. 
$r\vdash n$ is an irrep of $S_n$ and $s\vdash m$ is an irrep of $S_m$ so that $(r,s)$ is an irrep of the
$S_n\times S_m$ subgroup. 
In general $(r,s)$ will be subduced more than once.
The label $\alpha$ specifies which copy of the irrep we consider.
There is a significant simplification when $R$ has only two rows or columns: all irreps of $S_n\times S_m$ are subduced without multiplicity
so that the $\alpha$ index can be dropped.
To summarize, the restricted Schur polynomial is labeled by three Young diagrams, one for the $Z$s (which is a representation of $S_n$), 
one for the impurities (which is a representation of $S_m$) and one for the
``composition'' (which is any representation of $S_{n+m}$ that subduces $(r,s)$ when restricted
to the $S_n\times S_m$ subgroup). A comment on notation is in order. 
We will sometimes need to refer to two representations of the same group in a single equation.
In this case, 
the letters $r,t$ will be used to denote representations of $S_n$,
the letters $s,u$ will be used to denote representations of $S_m$ and
the letters $R,T$ will be used to denote representations of $S_{n+m}$.

Another ingredient appearing in (\ref{restschur}) is the restricted trace ${\rm Tr}_{(r,s)\alpha\beta}\left(\Gamma^{(R)}(\sigma)\right)$. The 
restricted trace is a ``trace'' over the elements of a subspace whose column index belongs to $S_n\times S_m$ irrep $(r,s)\alpha$ and
whose row index belongs to $S_n\times S_m$ irrep $(r,s)\beta$\footnote{We put the word ``trace'' in inverted commas because if $\alpha\ne\beta$
one is not even summing over diagonal matrix elements!}. To concretely see how this works it is useful to review the
method of \cite{Koch:2011hb,mn} which decomposes an $S_{n+m}$ irrep into $S_n\times S_m$ irreps.
Recall that standard tableaux are labelings of the Young diagram with integers $1$ to $m + n$ that are strictly decreasing down 
the columns and along the rows.
The standard tableaux provide a basis of states for irrep $R$ of $S_{n+m}$. 
Following \cite{Koch:2011hb,mn} we perform the reduction from $S_{n+m}$ to $S_n$ by introducing partially labeled Young tableaux,
which have $m$ boxes labeled, and the remaining $n$ boxes unlabeled.
These partially labeled Young tableaux stand for a collection of states. 
The integers $1$ to $m$ are in fixed locations and the integers $m + 1,...,m + n$ are then distributed in all possible locations thereby
recovering collections of standard tableaux.
Each such collection is a complete irrep of $S_n$.  
The unlabeled boxes determine the irrep $r$ of $S_n$. 
Our task is now to combine the partially labeled Young tableaux with $r$ fixed, into good irreps $s$ of $S_m$. 
An irrep $s$ can occur with some multiplicity. 
The labels $\alpha,\beta$ run over this multiplicity. 
Concretely we can write the restricted trace as
\bea
  {\rm Tr}_{(r,s)\alpha\beta}\left(\Gamma^{(R)}(\sigma)\right)={\rm Tr}_R \left(P_{R,(r,s)\alpha\beta}\Gamma^{(R)}(\sigma)\right)
\eea
where
\bea
  P_{R,(r,s)\alpha\beta} = {\bf 1}_r\otimes \,\sum_{i=1}^{d_s}\, |s\,\alpha\, ;\, i\rangle\langle s\,\beta\, ;\, i|
  \label{projector}
\eea
Here $i$ is a state label for the irrep $s$, $d_s$ is the dimension of irrep $s$ 
and ${\bf 1}_r$ is the projector inside the carrier space of $R$ onto irrep $r$.
By definition, ${\bf 1}_r$ gives 1 when acting on a partially labeled Young tableaux whose unlabeled boxes have shape $r$ and
gives zero otherwise.
To proceed further, we need to construct the factor
$$
\sum_{i=1}^{d_s}\, |s\,\alpha\, ;\, i\rangle\langle s\,\beta\, ;\, i|
$$
appearing in (\ref{projector}).
The construction of this factor developed in \cite{Koch:2011hb,mn} applies when the corners of Young diagram $R$ are well separated.
Each box in the Young diagram can be assigned a factor; the box in row $j$ and column $i$ has a factor $N+i-j$.
We consider only operators labeled by $R$ that have all rows of different length.
The right most box of each row therefore defines a corner on the right hand side of the Young diagram.
These corners of the Young diagram are well separated when the difference between the factors of these right most boxes are large. 
In our case, these differences for all corners on the right hand side of the Young diagram, go to infinity as we take the large $N$ limit.
We will call this the {\it displaced corners approximation}. 
As soon as the length of any pair of rows becomes similar, the displaced corners approximation no longer applies. 
In this limit \cite{Koch:2011hb,mn} have shown that the factor (\ref{projector}) can be constructed using simple Unitary group
representation theory.

Given this progress on diagonalizing the free field inner product it is natural to turn next to the spectrum of the 
dilatation operator in these non-planar large $N$ limits.
Initial numerical studies showed, remarkably, that the spectrum of the dilatation operator is that of a set of 
decoupled oscillators.
Early studies computed the exact action of the dilatation operator and then took the large $N$ limit as a final step.
These computations are quite involved and it is not easy to obtain general results.
Indeed, \cite{Koch:2010gp} focused on two impurities while \cite{VinceKate} considered 3 or 4 impurities.
By working in the displaced corners approximation, \cite{Carlson:2011hy} was able to directly implement the simplifications of the large $N$ 
limit allowing the computation of results for an arbitrary number of impurities but under the constraint that $R,r,s$ have at most two
columns or rows.
This was then extended beyond the su$(2)$ sector in \cite{Koch:2011jk} and to an arbitrary number of rows in \cite{Koch:2011hb}.
This extension used a novel Schur-Weyl duality \cite{Koch:2011hb,mn} that emerges at large $N$ in the displaced corners approximation.
Using this novel Schur-Weyl duality, the states $|s\,\mu_1\, ;\, i\rangle$ appearing in (\ref{projector}) are 
states of a U$(p)$ representation where $p$ is the number of rows or columns of the restricted Schur polynomial.
This allows us to trade the pair $s\,\mu_1$ for a Gelfand-Tsetlin pattern if we wish.
These results have a direct application to the sl$(2)$ sector\cite{deMelloKoch:2011vn}.
In the displaced corners approximation the action of the dilatation operator has an interesting structure.
The eigenproblem of the anomalous dimensions factors into a product of two problems, one for the $Z$s involving
Young diagram $r$ and one for the $Y$s involving Young diagram $s$.
In \cite{Koch:2011hb}, based on numerical results, a conjecture for the solution to the eigenproblem involving the $s$ label was given. 
This conjecture has now been proven in \cite{DCI}.
The starting point of \cite{DCI} is a proof that the number of excited giant graviton states
as constrained by the Gauss Law, matches the number of restricted Schur polynomials in the gauge theory. 
The proof proceeds by associating excited giant graviton states to elements of a double coset involving permutation groups. 
Making heavy use of the ideas and methods of \cite{BHR1,BHR2},
Fourier transformation on the double coset suggests an ansatz for the operators of a good scaling dimension. 
The operators obtained in this way, denoted $O_{R,r}(\sigma)$, are labeled by an element of a double coset $\sigma$ and by the Young
diagrams $R$ and $r$.
In \cite{DCI} it was proven that this ansatz indeed provides the conjectured diagonalization. 
Further, since the double coset structure is determined entirely by the Gauss Law which holds at all loops, 
these results suggest that the operators constructed in \cite{DCI}, may be relevant at higher loops.
This is an issue we will manage to probe in this article.
The eigenproblem on the $r$ label has been considered in \cite{gs}.
It is written in terms of a difference operator. 
The basic observation of \cite{gs} is to realize that this difference operator is an element of the Lie algebra of 
$U(p)$ when $r$ has $p$ rows or columns.
Exploiting this insight \cite{gs} argued that the eigenproblem on the $r$ label is related to a system of $p$ particles in a 
line with 2-body harmonic oscillator interactions.

We can now give the set of questions that motivated this study. 
As just discussed, the dilatation operator of ${\cal N}=4$ super Yang-Mills theory is integrable in the large $N$ displaced
corners approximation of the su$(2)$ sector at one loop\cite{Carlson:2011hy,Koch:2011hb,DCI}. 
The first question we wish to address is
\begin{itemize}
\item[1] {\bf Is the dilatation operator integrable in the large $N$ displaced corners approximation at higher loops?}
\end{itemize}
Although we are not able to give a complete answer to this question, we will test integrability at two loops. 
We can sharpen the above question. 
As described above, the action of the dilatation operator factorizes into an action on the Young diagram associated 
with the $Z$s and an action on the Young diagram associated with the $Y$s.
The eigenproblem associated with the $Y$s appears (see \cite{DCI}) to be determined by the Gauss Law constraint, 
which should hold at all higher loops. This motivates the question  
\begin{itemize}
\item[2] {\bf Do the $O_{R,r}(\sigma)$ of \cite{DCI} continue to solve the $Y$ eigenproblem at higher loops?}
\end{itemize}
The $Z$ eigenproblem was solved in \cite{gs} by mapping it to a system of $p$ particles in a line with 2-body harmonic 
oscillator interactions. The basic observation was to show that the operator to be diagonalized is an element of the Lie algebra 
of $U(p)$ when $r$ has $p$ rows or columns. Our third question is
\begin{itemize}
\item[3] {\bf Can the two loop $Z$ eigenproblem be mapped to a system of $p$ particles, again using the Lie algebra of $U(p)$?}
\end{itemize}
The one loop spectrum of anomalous dimensions has some interesting features. One would have expected the eigenvalues of the one
loop dilatation operator to be a function of the 't Hooft coupling. We find they are given by an integer times $g_{YM}^2$. It is
not completely clear how this should be interpreted. By computing the two loop correction to the anomalous dimension and requiring
that it is small compared to the leading term, we hope to
gain insight into both the interpretation of our results and in the precise limit that should be taken to get a sensible
perturbative expansion. This motivates our fourth question
\begin{itemize}
\item[4] {\bf Does the two loop correction to the anomalous dimension determine the precise limit that should be taken to get a sensible
perturbative expansion?}
\end{itemize}
These questions are all answered in the discussion section. We will find that this limit of the theory continues to be integrable at two loops, that
the one loop operators with a good scaling dimension are not modified at two loops and finally, that our perturbative expansion is
sensible in the conventional 't Hooft limit. 

This paper is organized as follows:
In the next section we will compute the action of the two loop dilatation operator in the large $N$ limit. 
We do this under the assumption that the number of $Z$s (which we denote by $n$) is much larger than the 
number of $Y$s (which we denote by $m$).
The condition $m\ll n$ is needed to ensure that the displaced corners approximation is justified.
The result of this computation is given in (\ref{twoloopdil}).
In section 3 we diagonalize the dilatation operator and obtain the spectrum of anomalous dimensions to two loops.
Section 4 is used to discuss these results and their relevance for the AdS/CFT correspondence.

\section{Two Loop Dilatation Operator}\label{ttwwoo}

Our goal is to evaluate the action of the two loop dilatation operator\cite{Beisert:2003tq}
\bea
  D_4 =&-&2 g^2:{\rm Tr}\left(\left[ \left[Y,Z\right],{\partial\over\partial Z}\right]
       \left[ \left[{\partial\over\partial Y},{\partial\over\partial Z}\right], Z\right]\right):\cr
       &-&2 g^2:{\rm Tr}\left(\left[ \left[Y,Z\right],{\partial\over\partial Y}\right]
       \left[ \left[{\partial\over\partial Y},{\partial\over\partial Z}\right], Y\right]\right):\cr
       &-&2 g^2:{\rm Tr}\left(\left[ \left[Y,Z\right],T^a \right]
        \left[ \left[{\partial\over\partial Y},{\partial\over\partial Z}\right], T^a \right]\right):
\label{dilop}
\eea
\bea
g={g_{YM}^2\over 16\pi^2}
\eea
on restricted Schur polynomials.
The normal ordering symbols here indicate that derivatives within the normal ordering
symbols do not act on fields inside the normal ordering symbols.
For the operators we study, $n\gg m$ so that only the first term in $D_4$ will contribute.
We have in mind a systematic expansion in two parameters: ${1\over N}$ and ${m\over n}$.
In Appenidx \ref{msmall} we show that keeping only the first term in $D_4$ corresponds to the computation
of the leading term in this double expansion.
The evaluation of the action of the one loop dilatation operator was carried out in \cite{VinceKate}.
The two loop computation uses many of the same techniques but there are a number of subtle
points that must be treated correctly. The computation can be split
into the evaluation of two types of terms, one having all derivatives adjacent to each other
(for example ${\rm Tr} (ZYZ\partial_Z \partial_Y \partial_Z)$) and one in which only two of the derivatives
are adjacent (for example ${\rm Tr} (YZ\partial_Z Z\partial_Y \partial_Z )$). We will deal with
an example of each term paying special attention to points that must be treated with care.

{\vskip 0.5cm}

\noindent
{\it First Term:} Start by allowing the derivatives to act on the restricted Schur polynomial
\bea
  {\rm Tr} (ZYZ\partial_Z \partial_Y \partial_Z)&&\!\!\!\!\!\!\!\!\! \chi_{R,(r,s)\alpha\beta}(Z,Y)
  = {mn(n-1)\over n! m!}\sum_{\psi\in S_{n+m}}{\rm Tr}_{(r,s)\alpha\beta}(\Gamma^{(R)}(\hbox{{\small $(1,m+2)\psi (m+1,m+2)$}}))\cr
&&\!\!\!\!\!\! \times \delta^{i_1}_{i_{\psi(1)}} Y^{i_2}_{i_{\psi(2)}}\cdots Y^{i_m}_{i_{\psi(m)}}(ZYZ)^{i_{m+1}}_{i_{\psi(m+1)}}
    \delta^{i_{m+2}}_{i_{\psi(m+2)}}Z^{i_{m+3}}_{i_{\psi(m+3)}}\cdots Z^{i_{m+n}}_{i_{\psi(m+n)}}
\eea
The two delta functions will reduce the sum over $S_{n+m}$ to a sum over an $S_{n+m-2}$ subgroup. This sum is most easily 
evaluated using the reduction rule of \cite{de Mello Koch:2004ws,de Mello Koch:2007uu}. The reduction rule rewrites the 
sum over $S_{n+m}$ as a sum over $S_{n+m-2}$ and its cosets. This is most easily done by making use of Jucys-Murphy elements 
whose action is easily evaluated. To employ the same strategy in the current computation, the action of the Jucys-Murphy 
element will only be the simple one if we swap the delta function from slot $m+2$ to slot $2$. This gives
\bea
&&\!\!\!\!\!\! {mn(n-1)\over n! m!}\sum_{\psi\in S_{n+m-2}}{\rm Tr}_{(r,s)\alpha\beta}(\Gamma^{(R)}(
\hbox{\small{$(1,m+2)(2,m+2)\psi(2,m+2) \hat{C} (m+1,m+2)$}}))\cr
&&\!\!\!\!\!\!\quad \times Y^{i_3}_{i_{\psi(3)}}\cdots Y^{i_m}_{i_{\psi(m)}}(ZYZ)^{i_{m+1}}_{i_{\psi(m+1)}}
    Y^{i_{m+2}}_{i_{\psi(m+2)}}Z^{i_{m+3}}_{i_{\psi(m+3)}}\cdots Z^{i_{m+n}}_{i_{\psi(m+n)}}
\eea
where $\hat{C}=(N+J_2)(N+J_3)$ with $J_i$ a Jucys-Murphy element
\bea
   J_i =\sum_{k=i}^{n+m}\, (i-1,k)
\eea
Since we sum over the $S_{n+m-2}$ subgroup, we can decompose $R\vdash m+n$ into a direct sum of terms which involve the irreps 
$R''\vdash m+n-2$ of the subgroup\footnote{In general if $R$ denotes a Young diagram, then $R'$ denotes a Young diagram that can
be obtained from $R$ by removing one box, $R''$ denotes a Young diagram that can be obtained from $R$ by removing two boxes
etc.}. As usual \cite{de Mello Koch:2004ws,de Mello Koch:2007uu}, for each term in the sum, $\hat{C}$ 
is equal to the product of the factors of the boxes that must be removed from $R$ to obtain $R''$.
To rewrite the result in terms of restricted Schur polynomials, note that
\bea
Y^{i_3}_{i_{\psi(3)}}&&\!\!\!\!\cdots Y^{i_m}_{i_{\psi(m)}}(ZYZ)^{i_{m+1}}_{i_{\psi(m+1)}}
Y^{i_{m+2}}_{i_{\psi(m+2)}}Z^{i_{m+3}}_{i_{\psi(m+3)}}\cdots Z^{i_{m+n}}_{i_{\psi(m+n)}}\cr
&&\!\!\!\!\! ={\rm Tr}\left({\small \psi (2,m+1,1)} Y\otimes Z\otimes Y^{\otimes\, m-2}\otimes Z\otimes Y\otimes Z^{\otimes \, n-2}\right)\cr
&&\!\!\!\!\! ={\rm Tr}\left({\small (2,m+2)\psi (2,m+1,1)(2,m+2)} Y^{\otimes\, m}\otimes Z^{\otimes \, n}\right)
\eea
and make use of the identity\cite{Bhattacharyya:2008rc}
\bea
{\rm Tr}(\sigma Z^{\otimes\, n}\otimes Y^{\otimes\, m}) = \sum_{T,(t,u)\beta\alpha}
{d_T n! m!\over d_t d_u (n+m)!}
{\rm Tr}_{(t,u)\alpha\beta}(\Gamma^{(T)}(\sigma^{-1}))
\chi_{T,(t,u)\beta\alpha}(Z,Y)
\eea
After this rewriting the sum over $S_{n+m-2}$ can be carried out using the fundamental orthogonality relation.
The result is
\bea
\nonumber
&&\qquad \sum_{T,(t,u)\gamma\delta} \sum_{R'',T''}{d_T n(n-1)m\over d_t d_u d_{R''}(n+m)(n+m-1)}c_{RR'}c_{R'R''}
\,\,\,\chi_{T,(t,u)\gamma\delta}(Z,Y)\cr
\nonumber
&&\!\!\!\!\!\!\!\!\!\!\!\!
   \times{\rm Tr}(I_{T'' R''}\hbox{\small{$(2,m+2,m+1)P_{R,(r,s)\alpha\beta}(1,m+2,2)$}}I_{R'' T''}
         \hbox{\small{$(2,m+2)P_{T,(t,u)\delta\gamma}(m+2,2,1,m+1)$}})
\eea
The intertwiner $I_{R''\, T''}$ is a map (see Appendix  D of \cite{Koch:2011hb} for details on its properties) 
from irrep $R''$ to irrep $T''$.  It  is only non-zero if $R''$ and $T''$ have the same shape.
Thus, to get a non-zero result $R$ and $T$ must differ at most, by the placement of two boxes.
We make further comments relevant for this trace before equation (\ref{talkedabout}) below.

{\vskip 0.5cm}

\noindent
{\it Second Term:} Evaluation of the second term is very similar. 
In this case however, taking the derivatives produces a single delta function, 
which will reduce the sum over $S_{n+m}$ to a sum over $S_{n+m-1}$.
The delta function should be in slot 1.
The reader wanting to check an example may find it useful to verify that
\bea
\nonumber
  :{\rm Tr} (YZ\partial_Z Z \partial_Y \partial_Z):&&\!\!\!\!\!\!\!\! \chi_{R,(r,s)\alpha\beta}(Z,Y)
  =\sum_{T,(t,u)\gamma\delta} \sum_{R',T'}{d_T n(n-1)m\over d_t d_u d_{R'}(n+m)}c_{RR'}\cr
&&\!\!\!\!\!\!\!\!\!\!\!\!\!\!\!\!\!\!\!\!\!\!\!\!\times{\rm Tr}(I_{T' R'}\hbox{\small{$(1,m+2,m+1)P_{R,(r,s)\alpha\beta}$}}I_{R' T'}
         \hbox{\small{$(1,m+1)P_{T,(t,u)\delta\gamma}$}})\chi_{T,(t,u)\gamma\delta}(Z,Y)
\label{talkedabout}
\eea
The intertwiner $I_{R'\, T'}$ is a map from irrep $R'$ to irrep $T'$.  It  is only non-zero if $R'$ and $T'$ have the same shape.
Thus, to get a non-zero result $R$ and $T$ must differ at most, by the placement of a single box.
It is perhaps useful to spell out explicitely the meaning of the trace above. 
The above trace is taken over the reducible $S_{n+m}$ representation $R\oplus T$. In addition, the projectors within the trace allow 
us to rewrite the permutations appearing in the trace as
\bea
{\rm Tr}\left(I_{T' R'}\hbox{\small{$\Gamma^{(R)}\Big((1,m+2,m+1)\Big)P_{R,(r,s)\alpha\beta}$}}I_{R' T'}
         \hbox{\small{$\Gamma^{(T)}\Big((1,m+1)\Big)P_{T,(t,u)\delta\gamma}$}}\right)
\eea

{\vskip 0.5cm}

The final result for the action of the dilatation operator is (this includes only the first term in 
(\ref{dilop}) since $n\gg m$)
\bea
   D_4\chi_{R,(r,s)\alpha\beta}(Z,Y)&&\!\!\!\!\!=-2g^2\sum_{T,(t,u)\gamma\delta}\sum_{R'\, T'}
               {d_T n(n-1)m c_{RR'}\over d_t d_u d_{R'}(n+m)}M^{(b)}_{R,(r,s)\alpha\beta\,\, T,(t,u)\gamma\delta}
               \chi_{T,(t,u)\delta\gamma}(Z,Y)\cr
\nonumber
&&\!\!\!\!\!\!\!\!\!\!\!\!\!\!\!\!\!\!\!\!\!\!\!\!\!\!\!\!-2g^2\sum_{T,(t,u)\gamma\delta}\sum_{R''\, T''}
               {d_T n(n-1)m c_{RR'}c_{R'R''}\over d_t d_u d_{R''}(n+m)(n+m-1)}M^{(a)}_{R,(r,s)\alpha\beta\,\, T,(t,u)\gamma\delta}
               \chi_{T,(t,u)\delta\gamma}(Z,Y)
\eea
where
\bea
&&\!\!\!\!\!\! M^{(a)}_{R,(r,s)\alpha\beta\,\, T,(t,u)\gamma\delta}=
     {\rm Tr}\left( I_{T''R''}\hbox{\small{$(2,m+2)P_{R,(r,s)\alpha\beta}
       C_1$}}I_{R''T''}\hbox{\small{$(2,m+2)P_{T,(t,u)\gamma\delta}C_1$}}\right)\cr
       \cr
     &&+ {\rm Tr}\left( I_{T''R''}\hbox{\small{$C_2 P_{R,(r,s)\alpha\beta}
       (2,m+2)$}}I_{R''T''}\hbox{\small{$C_2 P_{T,(t,u)\gamma\delta}(2,m+2)$}}\right)
\label{ma}
\eea
\bea
C_1=\big[ (m+2,2,1),(1,m+1)\big]\qquad C_2=-C_1^T = \big[ (m+2,1,2),(1,m+1)\big]
\eea
and
\bea
M^{(b)}_{R,(r,s)\alpha\beta\,\, T,(t,u)\gamma\delta}=
       {\rm Tr}\left( \hbox{\small{$I_{T'R'} C_3 I_{R'T'}\big[ (1,m+1) ,P_{T,(t,u)\gamma\delta}\big]$}}\right)\cr
    +  {\rm Tr}\left( \hbox{\small{$I_{T'R'} C_4 I_{R'T'}\big[ (1,m+1) ,P_{T,(t,u)\gamma\delta}\big]$}}\right)
\label{mb}
\eea
\bea
  C_3= \big[(1,m+2,m+1), P_{R,(r,s)\alpha\beta}\big]\qquad C_4 = \big[(1,m+1,m+2), P_{R,(r,s)\alpha\beta}\big]
\eea
This formula is correct to all orders in $1/N$. 
Denote the number of rows in the Young diagram $R$ labeling the restricted Schur polynomial by $p$.
This implies that, since $R$ subduces $S_n\times S_m$ representation $(r,s)$ and $n\gg m$ that $r$ has $p$ rows and
$s$ has at most $p$ rows. 
Now we will make use of the displaced corners approximation.
To see how this works, recall that to subduce $r\vdash n$ from $R\vdash m+n$ we remove $m$ boxes from $R$.
Each removed box is associated with a vector in a $p$ dimensional vector space $V_p$.
Thus, the $m$ removed boxes associated with the $Y$s thus define a vector in $V_p^{\otimes\, m}$.
In the displaced corners approximation, the trace over $R\oplus T$ factorizes into a trace over $r\oplus t$ and a trace over $V_p^{\otimes m}$. 
The structure of the projector (\ref{projector}) makes it clear that the bulk of the work is in evaluating the trace over $V_p^{\otimes m}$. 
This trace can be evaluated using the methods developed in \cite{Koch:2011hb}.
Introduce a basis for the fundamental representation of the Lie algebra u$(p)$ given by $(E_{ij})_{ab}=\delta_{ia}\delta_{jb}$.
Recall the product rule
\bea
  E_{ij}E_{kl} = \delta_{jk} E_{il}
\eea
which we use extensively below.
If a box is removed from row $i$ it is associated to a vector $v_i$ which is an eigenstate of $E_{ii}$ with eigenvalue 1.
The intertwining maps can be written in terms of the $E_{ij}$. For example, if we remove two boxes from
row $i$ of $R$ and two boxes from row $j$ of $T$, assuming that $R''$ and $T''$ have the same shape, we have
\bea
  I_{T''R''}=E^{(1)}_{ji}E^{(2)}_{ji}
  \label{int}
\eea
A big advantage of realizing the intertwiners in this way is that it is simple to evaluate the product of symmetric group
elements with the intertwiners. 
For example, using the identification (for background, see for example \cite{cmrii})
\bea
   (1,2,m+1)={\rm Tr}\big( E^{(1)}E^{(2)}E^{(m+1)}\big)
    \label{intt}
\eea
we easily find
\bea
  (1,2,m+1)I_{T''R''} \, = \, E^{(1)}_{kl}E^{(2)}_{lm}E^{(m+1)}_{mk}E^{(1)}_{ji}E^{(2)}_{ji} \, = \, E^{(1)}_{ki}E^{(2)}_{ji}E^{(m+1)}_{jk}
  \label{inttt}
\eea
This is now enough to evaluate the traces appearing in (\ref{ma}) and (\ref{mb}).

We will consider the action of the dilatation operator on normalized restricted Schur polynomials. 
The two point function for restricted Schur polynomials is \cite{Bhattacharyya:2008rb}
$$
\langle\chi_{R,(r,s)\alpha\beta}(Z,Y)\chi_{T,(t,u)\gamma\delta}(Z,Y)^\dagger\rangle =
\delta_{R,(r,s)\,T,(t,u)}\delta_{\alpha\gamma}\delta_{\beta\delta}\,
f_R {{\rm hooks}_R\over {\rm hooks}_{r}\, {\rm hooks}_s}
$$
where $f_R$ is the product of the factors in Young diagram $R$ and ${\rm hooks}_R$ is the product of the hook 
lengths of Young diagram $R$. The normalized operators are thus given by
$$
\chi_{R,(r,s)}(Z,Y)=\sqrt{f_R \, {\rm hooks}_R\over {\rm hooks}_r\, {\rm hooks}_s}O_{R,(r,s)}(Z,Y)\, .
$$
The components $m_i$ of the vector $\vec{m}(R)$ record the number of boxes removed from row $i$ of $R$ to produce $r$. 
In the su$(2)$ sector, both the one loop dilatation operator\cite{Koch:2011hb} and the two loop dilatation operator conserve $\vec{m}(R)$,
recorded in the factor $\delta_{\vec{m}(R)\vec{m}(T)}$ in (\ref{twoloopdil}) below.
In terms of these normalized operators the dilatation operator takes the form
\bea
   D_4 O_{R,(r,s)\mu_1\mu_2}= - 2 g^2\sum_{u\,\nu_1\,\nu_2}\delta_{\vec{m}(R)\vec{m}(T)}M^{(ij)}_{s\mu_1\mu_2\, ;\, u\nu_1\nu_2}
                   \left(\Delta^{(1)}_{ij}+\Delta^{(2)}_{ij}\right)O_{R,(r,u)\nu_1\nu_2}
   \label{twoloopdil}
\eea
where
\bea
M^{(ij)}_{s\mu_1\mu_2\, ;\, u\nu_1\nu_2}&&={m\over\sqrt{d_u d_s}}\left(
\left\langle \vec{m},s,\mu_2\, ;\, a |E^{(1)}_{ii}|\vec{m},u,\nu_2\, ;\, b\right\rangle
\left\langle \vec{m},u,\nu_1\, ;\, b |E^{(1)}_{jj}|\vec{m},s,\mu_2\, ;\, a\right\rangle\right.\cr
&&+\left.
\left\langle \vec{m},s,\mu_2\, ;\, a |E^{(1)}_{jj}|\vec{m},u,\nu_2\, ;\, b\right\rangle
\left\langle \vec{m},u,\nu_1\, ;\, b |E^{(1)}_{ii}|\vec{m},s,\mu_2\, ;\, a\right\rangle\right)
\label{GGbit}
\eea
To spell out the action of the operators $\Delta^{(1)}_{ij}$ and $\Delta^{(2)}_{ij}$ we will need a little more notation.
Denote the row lengths of $r$ by $r_i$. 
The Young diagram $r^+_{ij}$ is obtained by deleting a box from row $j$ and adding it to row $i$.
The Young diagram $r^-_{ij}$ is obtained by deleting a box from row $i$ and adding it to row $j$.
In terms of these Young diagrams define
\bea
\Delta^{0}_{ij}O_{R,(r,s)\mu_1\mu_2}=-(2N+r_i+r_j)O_{R,(r,s)\mu_1\mu_2}
\eea
\bea
\Delta^{+}_{ij}O_{R,(r,s)\mu_1\mu_2}=\sqrt{(N+r_i)(N+r_j)}\, O_{R^+_{ij},(r^+_{ij},s)\mu_1\mu_2}
\eea
\bea
\Delta^{-}_{ij}O_{R,(r,s)\mu_1\mu_2}=\sqrt{(N+r_i)(N+r_j)}\, O_{R^-_{ij},(r^-_{ij},s)\mu_1\mu_2}
\eea
We can now write
\bea
\Delta^{(1)}_{ij}= n(\Delta^+_{ij} +\Delta^0_{ij} + \Delta^-_{ij})
\eea
\bea
  \Delta^{(2)}_{ij}= (\Delta^+_{ij})^2 +\Delta^0_{ij}\Delta^+_{ij} +2\Delta^+_{ij}\Delta^-_{ij} +\Delta^0_{ij}\Delta^-_{ij} +(\Delta^-_{ij})^2
  \label{specstruc}
\eea
This completes the evaluation of the dilatation operator.

Our result for $\Delta^{(2)}_{ij}$ deserves a comment.
The intertwiners $I_{T''R''}$ appearing in (\ref{ma}) only force the shapes of $T$ and $R$ to agree when two boxes have been removed from each. 
One might imagine removing a box from rows $i,j$ of $R$ to obtain $R''$ and from rows $k,l$ of $T$ to obtain $T''$, implying that
in total four rows could participate. 
We see from $\Delta^{(2)}_{ij}$ that this is not the case - the mixing is much more constrained with only two rows participating.
We discuss this point further in Appendix \ref{intisspecial}.

\section{Spectrum}\label{Spectrum}

\begin{figure}[ht]%
\begin{center}
\includegraphics[width=10cm]{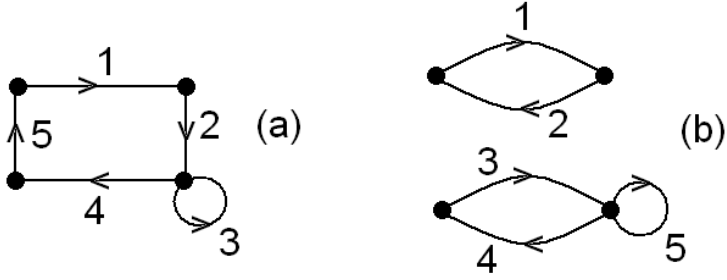}%
\caption{Two possible configurations for operators with $p=4$ and $m=5$.}%
\label{gaussexmpl}%
\end{center}
\end{figure}

\begin{figure}[ht]%
\begin{center}
\includegraphics[width=3cm]{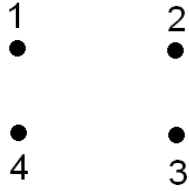}%
\caption{Labeling of the giant graviton branes.}%
\label{branelabel}%
\end{center}
\end{figure}

An interesting feature of the result (\ref{twoloopdil}) is that the action of the dilatation operator has factored into the
product of two actions: $\Delta^{(1)}_{ij}+\Delta^{(2)}_{ij}$ acts only on Young diagram $r$ i.e. on the $Z$s, while $M^{(ij)}_{s\mu_1\mu_2\, ;\, u\nu_1\nu_2}$
acts only on Young diagram $s$, i.e.  on the $Y$s.
This factored form, which also arises at one loop, 
implies that we can diagonalize on the $s\mu_1\mu_2 ; u\nu_1\nu_2$ and the $R,r ; T,t$ labels separately. 
The diagonalization on the $s\mu_1\mu_2 ; u\nu_1\nu_2$ labels is identical to the diagonalization problem which arises at one loop.
The solution was obtained analytically for two rows in \cite{Carlson:2011hy} and then in general in \cite{DCI}.
Each possible open string configuation consistent with the Gauss Law constraint can be identified with an element of a double coset.
A very natural basis of functions, constructed from representation theory, is suggested by Fourier transformation applied to this double coset. 
In this way \cite{DCI} constructed an explicit formula for the wavefunction which solves the $s\mu_1\mu_2 ; u\nu_1\nu_2$ diagonalization.
The resulting Gauss graph operators are labeled by elements of the double coset. 
The explicit solution obtained in \cite{DCI} is
\bea
  O_{R,r}(\sigma)
  ={|H|\over \sqrt{m!}}\sum_{j,k}\sum_{s\vdash m}\sum_{\mu_1,\mu_2}\sqrt{d_s}
  \Gamma^{(s)}_{jk}(\sigma )B^{s\to 1_H}_{j \mu_1}B^{s\to 1_H}_{k \mu_2} O_{R,(r,s)\mu_1\mu_2}
  \label{ggo}
\eea
where the group $H=S_{m_1}\times S_{m_2}\times \cdots\times S_{m_p}$ and the branching coefficients
$B^{s\to 1_H}_{j \mu_1}$ provide a resolution of the projector from irrep $s$ of $S_m$ onto the trivial
representation of $H$
\bea
{1\over |H|}\sum_{ \sigma \in H } \Gamma^{(s)}_{ik} ( \sigma )
=\sum_\mu B^{s \rightarrow 1_H}_{ i \mu}  B^{s \rightarrow 1_H}_{ k \mu}
\eea
The action of the dilatation operator on the Gauss graph operator is
\bea\label{actiondil} 
   D_4 O_{R,r}(\sigma) = -2g^2 \sum_{i<j}n_{ij} ( \sigma ) 
\left(\Delta_{ij}^{(1)}+\Delta_{ij}^{(2)}\right)\, O_{R,r}(\sigma)
\label{frrstdiag}
\eea
The numbers $n_{ij}(\sigma)$ can be read off of the element of the double coset $\sigma$. 
Each possible Gauss operator is given by a set of $m$ open strings stretched between $p$ different giant graviton branes. 
As an example, consider $p=4$ with $m=5$. Two possible configurations are shown in Figure \ref{gaussexmpl}. 
Label the open strings with integers from $1$ to $m=5$ for our example.
The double coset element can then be read straight from the open string configuration by recording
how the open strings are ordered as closed circuits in the graph are traversed. For the graphs shown,
(a) corresponds to $\sigma=(1245)(3)$ and (b) corresponds to $\sigma=(12)(34)(5)$. 
The numbers $n_{ij}(\sigma)$ tell us how many strings stetch between branes $i$ and $j$.
The branes themselves are numbered with integers from $1$ to $p$, as shown in Figure \ref{branelabel} for our example.
Thus, for (a) the non-zero $n_{ij}$ are $n_{12}=1$, $n_{23}=1$, $n_{34}=1$, and $n_{14}=1$.
Notice that we don't record strings that emanate and terminate on the same brane - string 3 in (a)
or string 5 in (b), in this example.
For (b) the non-zero $n_{ij}$ are $n_{12}=2$ and $n_{34}=2$. 
For the details, see \cite{DCI}.

To obtain the anomalous dimensions, inspection of (\ref{actiondil}) shows that we now have to solve the eigenproblem of
$\Delta_{ij}^{(1)}$ and $\Delta_{ij}^{(2)}$. 
The operator $\Delta_{ij}^{(1)}$ is simply a scaled version of the operator which plays a role in the one loop dilatation operator. 
The corresponding operator which participates at one loop was identified as an element of u$(p)$ \cite{gs}.
It is related to a system of $p$ particles in a line with 2-body harmonic oscillator interactions\cite{gs}.
The operator $\Delta_{ij}^{(2)}$ is new. 
Following \cite{gs}, a useful approach is to study the continuum limit of $\Delta_{ij}^{(1)}$ and $\Delta_{ij}^{(2)}$. 
Towards this end, introduce the variables
\bea
  y_{j}={r_{j+1}-r_1\over \sqrt{N+r_1}},\qquad j=1,2,3,...,p-1
\eea
which become continuous variables in the large $N$ limit. 
We have numbered rows so that $r_1<r_2<\cdots <r_p$.
In the continuum limit our Gauss graph operators become functions of $y_i$
\bea
  O_{R,r}(\sigma)\equiv O^{\vec{m}(R)}(\sigma , r_1, r_2,\cdots ,r_p )\to O^{\vec{m}(R)}(r_1,y_1,\cdots,y_{p-1})
\eea
Using the expansions
\bea
  \sqrt{(N+r_i)(N+r_j)}=N+r_1 \, + \, {y_i+y_j\over 2}\sqrt{N+r_1} \, - \, {(y_i-y_j)^2\over 8} \, + \, O\left({1\over \sqrt{N+r_1}}\right)
\eea
and
\bea
O^{\vec{m}(R)}&&\!\!\!\!\!\!\!\!(r_1,y_1,\cdots,y_i+{1\over\sqrt{N+r_1}},\cdots,y_j-{1\over\sqrt{N+r_1}},\cdots,y_{p-1})=
O^{\vec{m}(R)}(r_1,y_1,\cdots,y_{p-1})\cr
&&\!\!\!\!\!\!\!\!+{1\over\sqrt{N+r_1}}
\, {\partial\over\partial y_i}\, 
O^{\vec{m}(R)}(r_1,y_1,\cdots,y_{p-1})-{1\over\sqrt{N+r_1}}
\, {\partial\over\partial y_j}\, O^{\vec{m}(R)}(r_1,y_1,\cdots,y_{p-1})\cr
&&\!\!\!\!\!\!\!\! +{1\over 2(N+r_1)}
\left( {\partial\over\partial y_i} - {\partial\over\partial y_j}\right)^2\, O^{\vec{m}(R)}(r_1,y_1,\cdots,y_{p-1})
\eea
we find that in the continuum limit
\bea
   \Delta^{(1)}_{i+1\, j+1}O_{R,r}(\sigma)\to n\left[ \left( {\partial\over\partial y_i} - {\partial\over\partial y_j}\right)^2
                                      -{(y_i-y_j)^2 \over 4}\right]O^{\vec{m}(R)}(r_1,y_1,\cdots,y_{p-1})
\label{firstdelta}
\eea
\bea
   \Delta^{(1)}_{1\, i+1}O_{R,r}(\sigma)\to n\left[ \left( 2{\partial\over\partial y_i} + \sum_{j\ne i}{\partial\over\partial y_j}\right)^2
                                      -{y_i^2 \over 4}\right]O^{\vec{m}(R)}(r_1,y_1,\cdots,y_{p-1})
\label{seconddelta}
\eea
and
\bea
\hbox{\small{$\Delta^{(2)}_{i+1\, j+1}O_{R,r}(\sigma)\to 2(N+r_1)\left[ \left( {\partial\over\partial y_i} - {\partial\over\partial y_j}\right)^2
                                      -{(y_i-y_j)^2 \over 4}\right]O^{\vec{m}(R)}(r_1,y_1,\cdots,y_{p-1})$}}
\label{thirddelta}
\eea
\bea
   \Delta^{(2)}_{1\, i+1}O_{R,r}(\sigma)\to 2(N+r_1)\left[ \left( 2{\partial\over\partial y_i} + \sum_{j\ne i}{\partial\over\partial y_j}\right)^2
                                      -{y_i^2 \over 4}\right]O^{\vec{m}(R)}(r_1,y_1,\cdots,y_{p-1})
\label{fourthdelta}
\eea
Remarkably, in the continuum limit both $\Delta_{ij}^{(1)}$ and $\Delta_{ij}^{(2)}$ have reduced to scaled versions of exactly the same
operator that appears in the one loop problem. 
In the Appendix \ref{discrete} we argue for the same conclusion without taking a continuum limit.
This implies that the operators that have a good scaling dimension at one loop are uncorrected at two loops. 

It is now straight forward to obtain the two loop anomalous dimension for any operator of interest. 
An instructive and simple example is provided by $p=2$ 
with\footnote{The number $n_{12}^+$ counts the number of open strings stretching from giant graviton 1 to giant graviton 2;
the number $n_{12}^-$ counts the number of open strings stretching from giant graviton 2 to giant graviton 1.
The Gauss Law constraint forces $n_{12}^+=n_{12}^-$. See \cite{DCI} for more details.} $n_{12}=n_{12}^+ +n_{12}^-\ne 0$. 
In this case, the anomalous dimension $\gamma (g^2 )$ which is the eigenvalue of
\bea
   D=D_2+D_4
\eea
with\footnote{The normalization for both $D_2$ and $D_4$ follows \cite{Beisert:2003tq}. 
This normalization for $D_2$ is a factor of 2 larger than the normalization used in 
\cite{Koch:2010gp,VinceKate,Carlson:2011hy,deMelloKoch:2011vn,Koch:2011jk,Koch:2011hb,DCI,gs}.}
\bea
  D_2=-2g :{\rm Tr}\left(\left[Y,Z\right] \left[{\partial\over\partial Y},{\partial\over\partial Z}\right] \right):
\eea
and $D_4$ given in (\ref{dilop}), is
\bea
  \gamma = 16 q n_{12}^+ \big( g + (2N + 2r_1 + n) g^2\big)
\eea
\bea
  q = 0,1,2,..,M\quad n_{12}=0,1,2,...
  \label{limits}
\eea
where the upper cut off $M$ is itself a number of order $N$. Clearly, if the $g^2$ term is to be a small
correction to the leading term, we must hold $\lambda_g\equiv gN$ fixed, which corresponds to the usual 't Hooft limit.
The fact that the usual 't Hooft scaling leads to a sensible perturbative expansion in this sector of the theory
was already understood in \cite{Berenstein:2003ah}.
We then find
\bea
  \gamma = {16 q n_{12}\over N} \big( \lambda_g + (2 + 2{r_1\over N} + {n\over N} ) \lambda_g^2\big)
  \label{anomdim}
\eea
For a given open string plus giant system (i.e. a given $n_{12}$), in the large $N$ limit, $x={q\over N}$ varies continuously from
0 to $x={M\over N}$ implying that the spectrum of anomalous dimensions
\bea
  \gamma = 16 x n_{12} \big( \lambda_g + (2 + 2{r_1\over N} + {n\over N} ) \lambda_g^2\big)
  \label{spect}
\eea
is itself continuous. At finite $N$ this spectrum is discrete. Notice that since both $n$ and $r_1$ are of order $N$,
all three terms multiplying $\lambda_g^2$ in (\ref{spect}) are of the same size.
Note that the value for $\gamma$ (\ref{spect}) will recieve both ${1\over N}$ corrections and ${m\over n}$ corrections.

\section{Answers and Discussion}

We can now return to the questions we posed in the introduction:

\begin{itemize}
\item[1] {\bf Is the dilatation operator integrable in the large $N$ displaced corners approximation at higher loops?}
\end{itemize}
We don't know. 
We have however been able to argue that the dilatation operator is integrable in the large $N$ displaced 
corners approximation at two loops. 
This requires both sending $N\to\infty$ and keeping $m\ll n$ to ensure the validity of the displaced corners approximation.
At large $N$ with $m\sim n$ we do not know
how to compute the action of the dilatation operator and hence integrability in this
situation is an interesting open problem.
It seems reasonable to hope that integrability will persist in the large $N$ displaced 
corners approximation at higher loops.
\begin{itemize}
\item[2] {\bf Do the $O_{R,r}(\sigma)$ of \cite{DCI} continue to solve the $Y$ eigenproblem at higher loops?}
\end{itemize}
Yes, the Gauss graph operators do indeed solve the $Y$ eigenproblem at two loops. The $Y$ eigenproblem at two loops is identical
to the $Y$ eigenproblem at one loop, so that even the eigenvalues (given by $n_{ij}(\sigma)$ in (\ref{frrstdiag})$\,$) are unchanged.
The fact that the Gauss operators continue to solve the $Y$ eigenproblem does not depend sensitively on the coefficients of
the individual terms in the two loop dilatation operator (see Appendix C).
\begin{itemize}
\item[3] {\bf Can the two loop $Z$ eigenproblem be mapped to a system of $p$ particles, again using the Lie algebra of U$(p)$?}
\end{itemize}
We have indeed managed to map the $Z$ eigenproblem to the dynamics of $p$ particles (in the center of mass frame).
The two loop problem again has a very natural phrasing in terms of the Lie algebra of U$(p)$.
The one loop and two loop problems are different: they share the same eigenstates but have different eigenvalues.
The fact that the eigenstates are the same does depend sensitively on the coefficients of
the individual terms in the two loop dilatation operator (see Appendix C).
\begin{itemize}
\item[4] {\bf Does the two loop correction to the anomalous dimension determine the precise limit that should be taken to get a sensible
perturbative expansion?}
\end{itemize}
Yes - requiring that the two loop correction in (\ref{anomdim}) is small compared to the one loop term clearly implies that we should be
taking the standard 't Hooft limit. Our result then has an interesting consequence: at large $N$, $x = q/N$ becomes a continuous parameter
and we recover a continuous energy spectrum.
This is clearly related to \cite{de Wit:1988ct}.
At any finite $N$ the spectrum is discrete.

Our discussion has been developed for operators with a label $R$ that has $p$ long rows, which are dual to giant gravitons wrapping 
an S$^3\subset$AdS$_5$.
Operators labeled by an $R$ that has $p$ long columns are dual to giant gravitons wrapping an S$^3\subset$S$^5$.
The anomalous dimensions for these operators are easily obtained from our results in this article 
(see section D.6 of \cite{Koch:2011hb} for a discussion of this connnection). 
The $\Delta_{ij}^{(1)}$ for this case is obtained by replacing the $r_i\to -r_i$ and $r_j\to -r_j$ in (\ref{firstdelta}) and 
(\ref{seconddelta}), while $\Delta_{ij}^{(2)}$ for this case is obtained by replacing the $r_i\to -r_i$ and $r_j\to -r_j$ 
in (\ref{thirddelta}) and (\ref{fourthdelta}).
The result (\ref{frrstdiag}) is unchanged when written in terms of the new $\Delta^{(1)}_{ij}$ and $\Delta^{(2)}_{ij}$.

Finally, the fact that our operators are not corrected at two loops is remarkable. 
It is natural now to conjecture that they are in fact exact and will not be corrected at any higher loop.
This is somewhat reminiscent of the BMN operators\cite{Berenstein:2002jq}.
In that case it is possible to determine the exact anomalous dimensions as a function of the 't Hooft coupling 
$\lambda_g$\cite{Santambrogio:2002sb}.
Can we use similar methods to achieve this for the operators discussed in this article?

\section*{Acknowledgments}

We would like to thank Sanjaye Ramgoolam for helpful discussions.
This work is supported by the South African Research Chairs Initiative of the Department of 
Science and Technology and National Research Foundation.

\begin{appendix}

\section{$\Delta^{(2)}_{ij}$ as an element of u$(p)$}\label{discrete}

In this appendix we will argue that, at large $N$, the eigenstates of $\Delta^{(1)}_{ij}$ are also
eigenstates of $\Delta^{(2)}_{ij}$. 
We focus on the case that $p=2$.
Towards this end we will review relevant background from \cite{gs}.
Recall that in the fundamental representation of $u(N)$ the generators can be taken as
\bea
  (E_{kl})_{ab}=\delta_{ak}\delta_{bl}\qquad k,l,a,b=1,2,...,N
\eea
Introduce the operators (the labeling is such that $i>j$ i.e. $Q_{ij}$ is not defined if $i<j$)
\bea
  Q_{ij}={E_{ii}-E_{jj}\over 2}\qquad Q^+_{ij}=E_{ij},\qquad Q^-_{ij}=E_{ji}
  \label{algelements}
\eea
which obey the familiar algebra of angular momentum raising and lowering operators
\bea
  \big[ Q_{ij},Q_{ij}^+\big]   = Q^+_{ij}\qquad
  \big[ Q_{ij},Q_{ij}^-\big]   =-Q^-_{ij}\qquad
  \big[ Q^+_{ij},Q_{ij}^-\big] = 2Q_{ij}
\eea
Irreps of these su$(2)$ subalgebras can be labeled with the eigenvalue of
\bea
  L_{ij}^2\equiv Q_{ij}^-Q_{ij}^++Q_{ij}^2+Q_{ij}=
          Q_{ij}^+Q_{ij}^-+Q_{ij}^2-Q_{ij}
\eea
and states in the representation are labeled by the eigenvalue of $Q_{ij}$
\bea
  Q_{ij}|\lambda,\Lambda\rangle = \lambda |\lambda, \Lambda\rangle
  \qquad
  L_{ij}^2|\lambda,\Lambda\rangle = (\Lambda^2 + \Lambda) |\lambda, \Lambda\rangle
  \qquad
 -\Lambda\le\lambda\le\Lambda 
\eea
The restricted Schur polynomials can be identified with particular states in a definite
irrep. The reader may consult \cite{gs} for the details. 
Identifying the restricted Schur polynomials with states of a $U(p)$ representation allows us to write 
$\Delta^{(1)}_{ij}$ as a $u(p)$ valued operator
\bea
  \Delta^{(1)}_{ij}=n\left(\, -{1\over 2}(E_{ii}+E_{jj})+Q_{ij}^- +Q_{ij}^+\,\right)\equiv n\Delta_{ij}
\eea
Note that
\bea
  {\cal C}= E_{ii}+E_{jj}
\eea
commutes with all elements (\ref{algelements}) of the su$(2)$ algebra and hence defines a Casimir of this algebra. 
It is simply a constant times the identity in a given u$(p)$ irrep. 
It is not difficult to check \cite{gs} that $\Delta^{(1)}_{ij}$ defines a discrete oscillator with creation
operator given by
\bea
  A^\dagger = {1\over 2}(E_{ii}-E_{jj})+{1\over 2}E_{ij}-{1\over 2}E_{ji}\qquad \big[\Delta_{ij},A^\dagger\big]=-2A^\dagger
\eea
As pointed out in \cite{gs}, a correctly normalized creation operator is given by $a^\dagger$ with $A^\dagger =\sqrt{M}a^\dagger$,
where $M$ is introduced in (\ref{limits}).
It is straight forward to verify that $\Delta^{(2)}_{ij}$ is given by
\bea
  \Delta^{(2)}_{ij}=(Q^+)^2 - {{\cal C}\over 2}Q^+ +2Q^+Q^- -{{\cal C}\over 2}Q^- + (Q^-)^2 
\eea
and hence that
\bea
  \big[ \Delta^{(2)}_{ij},A^\dagger\big]=-4(\Delta_{ij}+{{\cal C}\over 4})A^\dagger -4Q^+ -4Q
  \label{aten}
\eea
In terms of a correctly normalized operator at large $N$ we have (the last two terms in (\ref{aten}) can be dropped in the limit)
\bea
  \big[ \Delta^{(2)}_{ij},a^\dagger\big]=-4(\Delta_{ij}+{{\cal C}\over 4})a^\dagger
\eea
There are two things worth noting at this point.
First, when acting in the basis of energy eigenstates, it is clear that $a^\dagger$ is indeed a creation operator
but, due to the appearance of $\Delta_{ij}$, with a ``state dependent frequency''.
Said differently, $a^\dagger$ continues to move us to higher eigenstates but the energies of these states are not equally spaced.
Second, we can show that this result is in perfect agreement with section \ref{Spectrum}.
To make a comparison with section \ref{Spectrum} we need to restrict attention to states for which the eigenvalue
of $\Delta_{ij}$ is finite, so that on this subspace we can replace $\Delta_{ij}+{{\cal C}\over 4}\to {{\cal C}\over 4}$.
Using the value for ${\cal C}$ computed in \cite{gs}, for any state of finite energy, we have
\bea
  \big[ \Delta^{(2)}_{ij},a^\dagger\big]=-2 \, ( 2N + 2r_1 ) \, a^\dagger
\eea
in perfect agreement with section \ref{Spectrum}.

\section{Simplifications of the $m\ll n$ limit}\label{msmall}

In this Appendix we will explain why keeping the first term in (\ref{dilop}) corresponds to computing the
leading term in a systematic expansion of the anomalous dimension in a series expansion in ${1\over N}$ and
${m\over n}$. Notice that the first term in (\ref{dilop}) contains two derivatives with respect to $Z$ and
one derivative with respect to $Y$, whilst the second term contains one derivative with respect to $Z$ and
two derivatives with respect to $Y$. Since the number of $Z$s (given by $n$) is much greater than the
number of $Y$s (given by $m$) we should expect the leading contribution to come from the first term in
(\ref{dilop}). In this Appendix we will demonstrate that this is indeed the case.

It is simplest to consider the expression (\ref{twoloopdil}). The factor $M^{(ij)}_{s\mu_1\mu_2\, ;\, u\nu_1\nu_2}$
includes 
\bea
\left\langle \vec{m},s,\mu_2\, ;\, a |E^{(1)}_{ii}|\vec{m},u,\nu_2\, ;\, b\right\rangle
\left\langle \vec{m},u,\nu_1\, ;\, b |E^{(1)}_{jj}|\vec{m},s,\mu_2\, ;\, a\right\rangle\cr
+
\left\langle \vec{m},s,\mu_2\, ;\, a |E^{(1)}_{jj}|\vec{m},u,\nu_2\, ;\, b\right\rangle
\left\langle \vec{m},u,\nu_1\, ;\, b |E^{(1)}_{ii}|\vec{m},s,\mu_2\, ;\, a\right\rangle
\label{thisterm}
\eea
which
involves traces over interwiners acting in $V^{\otimes m}$. It has no dependence on the representation $r$ 
of the $Z$s and hence, has no dependence on $n$. Thus, all $n$ dependence comes from the coefficient multiplying
the above term (\ref{thisterm}). We will therefore study the coefficient of this term. As a 
consequence of the fact that the first term in (\ref{dilop}) contains two derivatives with respect to $Z$ and
one derivative with respect to $Y$, this term will have a coefficient which includes the factor
\bea
{d_T n(n-1)m d_{r''}\over d_t d_u d_{R''}(n+m)(n+m-1)}
\eea
Recall that $r''$ is obtained by removing two boxes from $r$.
The factor of $d_{r''}$ is produced when we take two derivatives with respect to $Z$. In the limit
that $m\ll n$ we now find
\bea
{d_T n(n-1)m d_{r''}\over d_t d_u d_{R''}(n+m)(n+m-1)} =  {m \over d_u }\left[ 1 + O\left({m\over n}\right)\right]
\eea
For the second term in (\ref{dilop}), the corresponding factor is now
\bea
{d_T m(m-1)n d_{r'}\over d_t d_u d_{R''}(n+m)(n+m-1)}
\label{frstcoeff}
\eea
The Young diagram $r'$ is obtained by removing one box from $r$.
The factor of $d_{r'}$ is produced when we take a single derivative with respect to $Z$. In the limit
that $m\ll n$ we now find
\bea
{d_T m(m-1)n d_{r'}\over d_t d_u d_{R''}(n+m)(n+m-1)}=
{m(m-1)\over n d_u}\left[ 1 + O\left({m\over n}\right)\right]
\label{scndcoeff}
\eea
Notice that (\ref{scndcoeff}) is smaller than (\ref{frstcoeff}) by a factor of ${m\over n}$ as we expected. The second term in
(\ref{dilop}) will thus contribute at higher order in a systematic ${m\over n}$ expansion.

Finally, performing the sum over the Lie algebra index in the third term in (\ref{dilop}) gives a term that is identical to the
one loop dilatation operator, except that it is supressed by a power of $N$. Thus, it does not contribute to the leading order
in a large $N$ expansion.

Thus, to summarize, keeping only the first term in $D_4$ in (\ref{dilop}) corresponds to the computation
of the leading term in the double expansion in the parameters ${1\over N}$ and ${m\over n}$.

\section{On the action of the Dilatation Operator}\label{intisspecial}

In this Appendix we want to discuss how sensitively integrability depends on the coefficients of the individual terms appearing in $D_4$. 
We will start by making a few comments on the structure of $\Delta^{(2)}_{ij}$ that we obtained in (\ref{specstruc}).

Recall that we argued
\bea
\nonumber
  {\rm Tr} (ZYZ\partial_Z \partial_Y \partial_Z)\chi_{R,(r,s)\alpha\beta}(Z,Y)=
 \sum_{T,(t,u)\gamma\delta} \sum_{R'',T''}{d_T n(n-1)m\over d_t d_u d_{R''}(n+m)(n+m-1)}c_{RR'}c_{R'R''}
\,\,\,\chi_{T,(t,u)\gamma\delta}(Z,Y)\\
\nonumber
\times{\rm Tr}(I_{T'' R''}\hbox{\small{$(2,m+2,m+1)P_{R,(r,s)\alpha\beta}(1,m+2,2)$}}I_{R'' T''}
         \hbox{\small{$(2,m+2)P_{T,(t,u)\delta\gamma}(m+2,2,1,m+1)$}})
\eea
in section \ref{ttwwoo}. Focus on the trace appearing in the second line above. 
Assume that we obtain $R'$ from $R$ by dropping a box from row $i$ and that we obtain $R''$ from $R'$ by dropping a box from row $j$.
Further, assume that we obtain $T'$ from $T$ by dropping a box from row $k$ and that we obtain $T''$ from $T'$ by dropping a box from row $l$.
Clearly then, we are allowing four rows of the Young diagram to participate when the dilatation operator acts.
With these assumptions, we easily find (see (\ref{int}), (\ref{intt}) and (\ref{inttt}) as well as the discussion around these equations)
\bea
  I_{R''T''}=E^{(1)}_{ik}E^{(2)}_{jl}\qquad I_{T''R''}=E^{(1)}_{ki}E^{(2)}_{lj}
\eea
and
\bea
(m+2,2,1,m+1)I_{T'' R''}(2,m+2,m+1)=E^{(1)}_{li}E^{(m+1)}_{kj}
\label{inT}
\eea
\bea
(1,m+2,2)I_{R'' T''}(2,m+2)=E^{(1)}_{jk}E^{(m+2)}_{il}
\label{inTT}
\eea
In obtaining these results we have made heavy use of the simplifications in the action of the symmetric group that arise in the displaced
corners approximation. It is now a simple matter to find
\bea
{\rm Tr}(I_{T'' R''}\hbox{\small{$(2,m+2,m+1)P_{R,(r,s)\alpha\beta}(1,m+2,2)$}}I_{R'' T''}
         \hbox{\small{$(2,m+2)P_{T,(t,u)\delta\gamma}(m+2,2,1,m+1)$}})
\nonumber 
\eea
\bea
={\rm Tr}(E^{(1)}_{li}E^{(m+1)}_{kj}P_{R,(r,s)\alpha\beta}E^{(1)}_{jk}E^{(m+2)}_{il}P_{T,(t,u)\delta\gamma})
\eea
Since the projectors $P_{R,(r,s)\alpha\beta}$ and $P_{T,(t,u)\delta\gamma}$ have a trivial action on slots $m+1$
and $m+2$, the above result is only non-zero when $i=l$ and $k=j$ - so that only two rows participate.

This reduction from four possible rows participating to two rows participating is determined by (\ref{inT}) and (\ref{inTT}).
These equations are corrected when going beyond the displaced corners approximation and, in that case, all four rows do indeed
enter. For all of the terms appearing in the first line of $D_4$, we find this reduction to two rows for each term separately.
Further, we find that each trace is individualy proportional to $M^{(ij)}_{s\mu_1\mu_2\, ;\, u\nu_1\nu_2}$ defined in
(\ref{GGbit}). This implies that the answer to question 2 that we posed in the introduction is completely insensitive
to the precise coefficients of the terms appearing in $D_4$\footnote{If one includes the remaining (subleading) terms in $D_4$ that
we have discarded in the $m\ll n$ limit, the dilatation operator starts to mix different Gauss graph operators. This suggests
that the integrability we study here is a property of the large $N$ limit and of the displaced corners approximation 
(i.e. $m<<n$) and may not survive when subleading corrections are included.}.

At this point it is natural to ask if the reduction of the dilatation operator to a set of decoupled oscillators (and thus the
observed integrability) is likewise also insensitive to the detailed coefficients. 
We will see that this is not the case - the emergence of an oscillator
does depend sensitively on the precise values of the coefficients of the terms appearing in $D_4$.

Consider equation (\ref{specstruc}). Individual terms appearing in (\ref{specstruc}) can be traced back to
particular terms appearing in $D_4$. For example, 
the terms proportional to $(\Delta^+_{ij})^2$ and $(\Delta^-_{ij})^2$ come from the terms ${\rm Tr} (ZZY\partial_Z\partial_Z\partial_Y)$
and ${\rm Tr} (YZZ\partial_Y\partial_Z\partial_Z)$. Notice that these two terms are related by daggering. Similarly, the terms
$\Delta^0_{ij}\Delta^+_{ij}$ and $\Delta^0_{ij}\Delta^-_{ij}$ come from the terms ${\rm Tr} (ZYZ\partial_Z\partial_Z\partial_Y)$
${\rm Tr} (ZZY\partial_Z\partial_Y\partial_Z)$, ${\rm Tr} (ZYZ\partial_Y\partial_Z\partial_Z)$ and
${\rm Tr} (YZZ\partial_Z\partial_Y\partial_Z)$ which are again related by daggering.
Changing the relative weights of terms appearing in $D_4$ will change the relative weight of terms appearing in (\ref{specstruc}).

To explore the effect of these changed coefficients on integrability, imagine we assign coefficient $\alpha$ to the terms 
${\rm Tr} (ZZY\partial_Z\partial_Z\partial_Y)$ and ${\rm Tr} (YZZ\partial_Y\partial_Z\partial_Z)$ in $D_4$. We now find
$\Delta^{(2)}_{ij}$ is replaced by
\bea
  \Delta^{\alpha (2)}_{ij}= \alpha(\Delta^+_{ij})^2 +\Delta^0_{ij}\Delta^+_{ij} +2\Delta^+_{ij}\Delta^-_{ij} 
                    +\Delta^0_{ij}\Delta^-_{ij} +\alpha (\Delta^-_{ij})^2
\eea
It is straight forward to check, using the approach of \cite{gs} that this operator does not admit creation and annihilation
operators and hence does not define an oscillator. A very instructive way to get some insight into what is going on, is to
consider the continuum limit of section \ref{Spectrum}. We find
\bea
  \Delta^{\alpha (2)}_{ij} O_{R,r}(\sigma)\to 2N^2(\alpha -1)O_{R,r}(\sigma) + 2(r_i+r_j)N(\alpha-1)O_{R,r}(\sigma) +O(N)
  \label{deformeddelta}
\eea
Compare this to (\ref{thirddelta}) and (\ref{fourthdelta}). Even the scaling with $N$ of the eigenvalues of $\Delta^{\alpha (2)}_{ij}$
and $\Delta^{(2)}_{ij}$ disagree. Indeed, with $\alpha=1$ we have a delicate cancelation of the leading order terms - as we clearly
see in (\ref{deformeddelta}). It is the subleading terms that combine to produce an oscillator. Note that all of the terms in
(\ref{specstruc}) contribute at the leading order. Thus, the sensitive dependence we see on the coefficient of the terms 
${\rm Tr} (ZZY\partial_Z\partial_Z\partial_Y)$ and ${\rm Tr} (YZZ\partial_Y\partial_Z\partial_Z)$ extends to the other terms in $D_4$
too. 

This last point deserves explanation.
The terms in $\Delta^{(2)}_{ij}$ can be collected into three groups which are each hermittian: 
$(\Delta^+_{ij})^2 + (\Delta^-_{ij})^2$, $\Delta^0_{ij}\Delta^+_{ij} + \Delta^0_{ij}\Delta^-_{ij}$ and finally
$2\Delta^+_{ij}\Delta^-_{ij}$. The relative coefficients of the terms producing these pieces is fixed by hermitticity. For example
${\rm Tr} (ZZY\partial_Z\partial_Z\partial_Y)+\beta {\rm Tr} (YZZ\partial_Y\partial_Z\partial_Z)$ is only hermittian if
$\beta=1$ and in this case the terms sum to $(\Delta^+_{ij})^2 + (\Delta^-_{ij})^2$. The particular
coefficients of the terms that appear in $\Delta^{(2)}_{ij}$ ensure that when we take the continuum limit (i) the terms proportional
to $N^2$ cancel, (ii) the terms proportional to $(r_i+r_j)N$ cancel and (iii) the surviving terms 
sum to produce an operator that admits exactly the same creation and annihilation operators as the one loop dilatation operator does. 
The integrability we have studied here depends on a careful fine tuning of the terms appearing in $D_4$.

\end{appendix}

\end{document}